\documentclass[twocolumn,preprintnumbers,superscriptaddress,endnote,nofootinbib,prl]{revtex4}
\usepackage{graphicx}

\usepackage[hypertex]{hyperref}
\newcommand{\vev}[1]{\langle {#1} \rangle}
\newcommand{\lsim}{\lesssim}
\newcommand{\gsim}{\gtrsim}

\newcommand{\eq}[1]{Eq.~(\ref{#1})}

\newcommand{\ord}[1]{\mathcal{O}{(#1)}}
\newcommand{\beq}{\begin{equation}}
\newcommand{\eeq}{\end{equation}}
\newcommand{\eps}{\varepsilon}

\begin{document}

\pagestyle{plain}

\title{Charged Higgs Discovery in the {\boldmath $W$} plus ``Dark'' Vector Boson Decay Mode}

\author{Hooman Davoudiasl
\footnote{email: hooman@bnl.gov}
}
\author{William J. Marciano
\footnote{email: marciano@bnl.gov}
}

\affiliation{Department of Physics, Brookhaven National Laboratory,
Upton, NY 11973, USA}

\author{Raymundo Ramos
\footnote{email: raramos@email.wm.edu}
}
\author{Marc Sher
\footnote{email: mtsher@wm.edu}
}

\affiliation{High Energy Theory Group, College of William and Mary,
Williamsburg, Virginia 23187, USA}


\begin{abstract}

In Two Higgs doublet extensions of the Standard Model, flavor-changing neutral current constraints can be
addressed by introducing a $U(1)'$ gauge symmetry,
under which the Higgs doublets carry different charges.  That scenario implies
the presence of a $H^\pm W^\mp Z'$ vertex at tree level.
For the light ``dark" $Z$ model ($Z'=Z_d$) with
$m_{Z_d} < 10$~GeV, such a coupling leads to the dominant
decay mode $H^\pm \to W^\pm + Z_d$ (for $m_{H^\pm} \lsim 175$~GeV),
rather than the usual type I model decay
$H^\pm \to \tau^\pm \nu$, for a broad range of parameters.  We find that
current analyses do not place significant bounds on this scenario.  Over much of the
parameter space considered, the decay of a pair-produced $t$ (${\bar t}$)
into $H^+ \,b$ ($H^- \,{\bar b}$) provides the dominant $H^\pm$ production.
Analysis of available LHC data can likely cover significant
ranges of our parameters, if $Z_d \to \mu^+\mu^-$ has a branching ratio of $\sim 20\%$.
If the $Z_d$ decays mainly invisibly then probing the entire relevant parameter space
would likely require additional data from future LHC runs.  We briefly discuss the
phenomenology for $m_{H^\pm}\gsim 175$~GeV.

\end{abstract}
\maketitle

Discovery of the neutral Higgs scalar boson $h$, with
mass $m_h \simeq 126$~GeV \cite{HiggsATLAS,HiggsCMS}, completes the $SU(3)_c \times
SU(2)_L \times U(1)_Y$ Standard Model (SM) elementary particle spectrum. So far, the observed properties of the Higgs
support the minimal paradigm of electroweak symmetry breaking and mass generation via a single complex
fundamental $SU(2)_L$ Higgs doublet, $\phi$, with vacuum expectation value  $\vev{\phi} = v/\sqrt{2}$, where
$v \simeq 246$~GeV.  Nevertheless, outstanding questions, such as the nature of dark matter (DM), origin of baryogenesis, and disparity of fermion masses remain and suggest the likelihood of additional ``new physics" beyond
SM expectations.

Among the simplest extensions of the SM are: {\it (1)} Introduction of a second Higgs doublet $\phi_2$
with $\vev{\phi_2}=v_2/\sqrt{2}$ which in tandem with $\vev{\phi_1} = v_1/\sqrt{2}$ breaks the $SU(2)_L \times U(1)_Y$
symmetry (with $v= \sqrt{v_1^2 + v_2^2} \simeq 246$~GeV) and gives rise to a SM like $h$, as well as additional
physical scalars $H$, $A$, and $H^\pm$ (yet to be observed).  {\it (2)}  Appendage of an extra $U(1)'$
local gauge symmetry broken by a complex singlet scalar field $\phi_s$, which gives mass to
its associated spin-1 gauge boson $Z'$ and introduces an additional
scalar $H_s$.  Discovery of a $Z'$ or additional scalar particles would fundamentally
revolutionize particle physics.

In this paper, we consider a combination of the two ``new physics" scenarios mentioned above, which form a
natural complement to the SM.  Two Higgs doublets $\phi_1$ and $\phi_2$ with  different $U(1)'$ gauge charges are
introduced.  The $U(1)'$ symmetry is utilized \cite{Davoudiasl:2012ag} to prevent the occurrence of tree level
flavor-changing neutral currents \cite{Glashow:1976nt}.  A singlet
$\phi_s$ is also introduced to give mass to the $Z'$.  Here, $\phi_2$ and all other SM particles are
assumed to have zero $U(1)'$ charge (this matches the conventional definition, in which $\phi_2$ couples to the top quark \cite{HHG,Branco:2011iw}) and hence all SM fermions only couple to $\phi_2$ (which acts very much like
the conventional SM Higgs doublet).  However,
$\phi_1$ and $\phi_s$ $U(1)'$ charges are non-vanishing.  This type of scenario is classified
as a type I 2-Higgs  doublet model (2HDM).  To agree with constraints from $\Delta
M_{B_d}$ (see, for example, Ref.~\cite{Chen:2013kt}), requires $v_2\gsim v_1$

Extra $U(1)'$ 2HDMs have been considered in the past for heavy $Z'$ bosons \cite{heavyZ'}.
The novelty of our approach is that we assume the $Z'$ is relatively light;
$m_{Z'} \lsim 10$~GeV.   (Generalization of our results to higher masses
is straightforward.)
Such a scenario is often associated with ``dark"
photon models in which $U(1)' = U(1)_d$ and $Z' = \gamma_d$
(the dark photon), a gauge symmetry connected with generic DM
or a hidden sector that is otherwise decoupled from the SM.  Dark
photons which kinetically mix with ordinary photons, parametrized
by $\eps \ll 1$, and acquire small induced couplings to our world \cite{Holdom:1985ag}
have been suggested in response to various astrophysical puzzles
\cite{ArkaniHamed:2008qn}, as well as an explanation for the muon
anomalous magnetic moment ($g_\mu - 2$) discrepancy
between theory and experiment \cite{Fayet:1990wx,Pospelov:2008zw}.

With an extra doublet, $\phi_1$, carrying $SU(2)_L \times U(1)_Y$ as well as $U(1)_d$ charges,
the SM $Z$ and $\gamma_d$
will undergo mass mixing with a small mixing parameter $\eps_Z$,
constrained by electroweak phenomenology.  This setup results
in the observed $Z$ and a light ``dark" $Z = Z_d$ \cite{Davoudiasl:2012ag}, where
\beq
\eps_Z \simeq \frac{m_{Z_d}}{m_Z} \cos \beta \, \cos \beta_d\,.
\label{epsZ}
\eeq
One finds
\beq
m_Z \simeq \frac{g}{2 \cos \theta_W} \sqrt{v_1^2 + v_2^2} \simeq 91~{\rm GeV}
\label{mZ}\eeq
and
\beq
m_{Z_d} \simeq g_d\, Q_d \, \sqrt{v_d^2 + v_1^2} \lsim 10~{\rm GeV}\,,
\label{mZ}\eeq
where $g$ and $g_d$ are, respectively, $SU(2)_L$ and $U(1)_d$ gauge couplings, $Q_d $ is the dark
charge of $\phi_s$ and $\phi_1$; $\vev{\phi_s} = v_d/\sqrt{2}$.   The convention used here
\beq
\tan \beta = v_2/v_1 \quad \quad ; \quad \quad \tan \beta_d = v_d/v_1
\label{tan}
\eeq
is inverted relative to  Ref.~\cite{Davoudiasl:2012ag}, where the model was first introduced, in keeping with the more standard 2HDM convention \cite{HHG,Branco:2011iw}.

The ``dark" $Z$ gives rise to interesting particle physics
phenomenology such as ``dark" parity violation \cite{Davoudiasl:2012qa}, rare meson
decays $K\to \pi Z_d$ and $B\to K Z_d$ (phase space permitting) and even $h\to Z Z_d$.  Constraints from these phenomena lead (very roughly and rather strongly dependent on $m_{Z_d}$) to \cite{Davoudiasl:2012ag}
\beq
\cos\beta \, \cos \beta_d \lsim 10^{-2}\,.
\label{betaconst}
\eeq
One can always go to zero mixing via $v_1\to 0$
and essentially decouple the effect of $Z_d$.

Given the above considerations, for our subsequent discussion, we will allow
\beq
5 \lsim \tan \beta \lsim 20 \quad \quad ; \quad \quad 5 \lsim \tan\beta_d \lsim 20
\label{tanrange}
\eeq
and charged Higgs masses
\beq
100~{\rm GeV} \lsim m_{H^\pm} \lsim 175{\rm ~GeV}
\label{massrange}
\eeq
as acceptable and interesting  parts of parameter space.  Larger values of $\tan\beta$ and
$\tan\beta_d$ are allowed, but may require fine-tuning to avoid problems with
unitarity (see Appendix A of Ref. \cite{Branco:2011iw} for a discussion and references) and
perturbation theory \cite{Chen:2013rba}.
Later, we will briefly discuss aspects of phenomenology for $m_{H^\pm}\gsim 175$~GeV.
We will mainly consider $m_{Z_d} \lsim 10~{\rm GeV}$
(larger $m_{Z_d}$ values are, of course, possible, but would have phase space effects on the predictions below).
For ranges of $\tan\beta$ and $\tan\beta_d$ in \eq{tanrange}, we would typically expect
$v_d \gsim 100$~GeV and hence it is implicitly assumed that $g_d \times Q_d \lsim 0.1$.

To highlight the main new point of this paper, we note that for any number of $SU(2)_L\times U(1)_Y$ Higgs doublets and
singlets in the extended SM, there is no $H^\pm W^\mp Z$ vertex at tree level.  Similarly,
if all doublets had the same $U(1)'$ couplings there would be no tree level $H^\pm W^\mp Z'$ vertex.  However,
since in the ``dark" $Z$ model the doublets have different $U(1)_d$ charges,
$Q_d(\phi_2)=0$ and (for definiteness) $Q_d(\phi_1)=Q_d(\phi_s)=1$, there
will be a tree level $H^\pm W^\mp Z_d$ vertex
induced by the kinetic covariant derivative Lagrangian
term $|D_\mu \phi_1|^2$, with
\beq
D_\mu = \partial_\mu + i g \frac{\tau^a}{2} W^a_\mu + i \frac{g'}{2} B_\mu + i g_d Z_{d\mu}\,,
\label{Dmu}
\eeq
where $\tau^a$, $a=1,2,3$, are Pauli matrices, $W^a_\mu$ are $SU(2)_L$ gauge fields, and
$g'$ and $B_\mu$ are the coupling constant and gauge field of $U(1)_Y$, respectively.
One finds a $H^\pm W^\mp Z_d$ coupling given by
\beq
\pm i \,g \, m_{Z_d} \sin\beta \, \cos\beta_d\, g_{\mu\nu}.
\label{HWZd-coupling}
\eeq
The $m_{Z_d}$ dependence in (\ref{HWZd-coupling})
is cancelled by a $1/m_{Z_d}$ factor when the longitudinal $Z_d$ is involved.
Hence, the only real suppression in that case comes from $\cos\beta_d \lsim 0.2$.

The coupling in \eq{HWZd-coupling} leads to the $H^\pm$ decay rate (neglecting $m_{Z_d}$ phase
space effects)
\begin{eqnarray}
&~&\Gamma(H^\pm \to W^\pm Z_d)  =  \nonumber\\
&~& \frac{g^2}{64 \pi}\frac{m_{H^\pm}^3}{m_W^2}
(\sin\beta\, \cos\beta_d)^2 \left(1 - \frac{m_W^2}{m_{H^\pm}^2}\right)^3\,.
\label{HtoWZd-rate}
\end{eqnarray}
That is to be compared with the primary fermionic decay (for $m_{H^\pm}\lsim 175$~GeV)
\begin{eqnarray}
&~&\Gamma(H^\pm \to \tau^\pm \nu)  =  \nonumber\\
&~& \frac{g^2}{32 \pi} m_{H^\pm} \frac{m_\tau^2}{m_W^2}
\cot^2\beta \left(1 - \frac{m_\tau^2}{m_{H^\pm}^2}\right)^2\,,
\label{Htotaunu-rate}
\end{eqnarray}
which is suppressed by $\cot^2\beta$ for the type I 2HDM, as well as the small $m_\tau^2/m_W^2$ factor.

The result in \eq{HtoWZd-rate} can also be simply derived in the limit $m_{H^\pm} \gg
m_W, m_{Z_d}$ using the Goldstone boson equivalence theorem, from the $H^\pm G^\mp a^0$
coupling $\pm g  \sin\beta\cos\beta_d \, m_{H^\pm}^2/(2 m_W)$, where $G^\pm$ and $a^0$
correspond to the Goldstone bosons that become longitudinal components of
the $W^\pm$ and $Z_d$. That computation confirms \eq{HtoWZd-rate} and provides a nice
check on the normalization of our result.

For the ratio $R \equiv \Gamma(H^\pm \to W^\pm Z_d)/\Gamma(H^\pm \to \tau^\pm \nu)$, one finds
\beq
R \simeq \frac{1}{2} \frac{m_{H^\pm}^2}{m_\tau^2}
\frac{\sin^4\beta}{\cos^2\beta} \cos^2\beta_d \left(1 - \frac{m_W^2}{m_{H^\pm}^2}\right)^3\,.
\label{R}
\eeq
Over the parameter space in \eq{tanrange} and \eq{massrange},
$R\gsim 1$ and the decay channel $H^\pm \to W^\pm Z_d$ has a dominant branching ratio near 1,
{\it i.e.} nearly all $H^\pm$ produced will give rise to $W^\pm \,Z_d$ final states.  Here, we have
implicitly assumed that all neutral scalars present in our 2HDM framework are heavier
than $\Delta m=m_{H^\pm}-m_W^\pm$ and hence on-shell
decays of the type $H^\pm \to W^\pm + \text{scalar}$ are
kinematically forbidden.  Note that for the mass range (\ref{massrange}),
$\Delta m \lsim 95$~GeV, which can be readily accommodated for a general scalar potential, including
terms proportional to $\phi_s \phi_1^\dagger \phi_2$ \cite{heavyZ'}.
Hence, the significance of $H^\pm \to W^\pm Z_d$ in our framework is quite general.
(Ref.~\cite{Maitra:2014qea} considers $H^\pm \to W^+ +$ invisible scalar in a different model.
If our model contains light scalars, $H^\pm \to W^\pm +$ scalar can have
a significant branching fraction, potentially leading to a
$W^+W^- + 4\, Z_d$ signature, which has been discussed in Ref.~\cite{leesher}.  For a discussion
of $H^\pm \to W^+ +$ pseudo scalar, see Ref.~\cite{Dermisek:2013cxa}.)

Since $H^+ \to W^+ Z_d$ can be the dominant decay
for reasonable values of the parameters, we now focus on its detection prospects.
Detection strategies differ depending on whether the $Z_d$ decays visibly or invisibly.
Often, it is assumed that the $Z_d$ decays will be exclusively to
charged states if kinetic mixing dominates,
but could also have a non-negligible rate into neutrinos if mass-mixing is
dominant \cite{Davoudiasl:2012ag}.  However, if the $Z_d$ is the mediator
of a force among light dark matter particles its decays are then expected
to be largely into invisible dark matter states (this possibility could potentially
be further probed at low energy fixed target experiments \cite{darkbeam1,darkbeam2}).

The partial decay widths of the $Z_d$ are given, assuming  only kinetic and mass matrix mixing, by
\cite{Davoudiasl:2012ag}
\beq
\Gamma(Z_d \to f \bar f) \simeq \frac{N_C}{48\pi}
\left(\frac{\eps_Z\, g}{\cos\theta_W}\right)^2 \left(g'^2_{V f} + g_{A f}^2\right) m_{Z_d} ,
\eeq
where fermion masses have been ignored, $N_C = 3\ (1)$
for quarks (leptons), $g_{A f} \equiv -T_{3f}$,
and $g'_{V f} \equiv T_{3f} - 2Q_f [\sin^2\theta_W - (\eps/\eps_Z)\cos\theta_W\sin\theta_W]$.   For $m_{Z_d}$
in the range of interest here, fermion masses and hadronic contributions
can be significant.  Dark photons decay with
individual leptonic branching fractions of $10\%$-$40\%$ are expected over our $m_{Z_d}$ range.
The branching fractions for $Z_d$ will be different and have
dependence on $\eps/\eps_Z$.   In what follows, for concreteness, we will assume a branching
fraction into $\mu^+\mu^-$ of $20\%$.  There may also be a significant invisible
rate for decay into neutrinos.  We will next consider various production
modes for the charged Higgs states.

Single production from light fermion initial states is negligible, since
in a type I 2HDM in which  $\tan\beta$ is not small,
the $H^\pm$ couplings to light fermions are quite suppressed.  However, over a significant
portion of the ranges in \eq{tanrange} and \eq{massrange},
single production of $H^\pm$ from the decay of a pair produced top or anti-top 
dominates.  To see this, note that the ratio $r_t \equiv \Gamma(t\to H^+ \,b)/\Gamma(t\to W^+ \, b)$ is
given by \cite{HHG}
\beq
r_t \simeq \frac{(1 - \mu_{H^\pm}^2)^2}{(1 - \mu_W^2)
}\cot ^2\beta\,,
\label{rt}
\eeq
where $\mu_i \equiv  m_i/m_t$.  For example, if $m_{H^\pm} = 120$~GeV and $\tan\beta=5$, we have
$r_t\simeq 0.01$.  Given that the $t{\bar t}$ production cross section at the
8 TeV LHC is $\sim 200$~pb \cite{ttbar},
we find a single $H^+$ cross section, from $t\to H^+ \,b$,
of about 2~pb.  Hence, the single production cross section of a charged Higgs, from a $t{\bar t}$ pair,
is about 4~pb.  For $m_{H^\pm}=160$~GeV, $r_t \simeq 10^{-3}$, which
would yield $\sim 0.4$~pb.  We can compare these to other production modes.

The process $b\, g \to t \,H^-$ is a significant production mode for a single charged Higgs \cite{Klasen:2012wq,Gong:2012ru} at the LHC.
For example, from Ref.~\cite{Gong:2012ru} (after adapting their results to a type I 2HDM) we 
find that for $\tan\beta = 10$ (typical of our parameter space),
the single production of a $\sim 160$~GeV
charged Higgs has a cross section of $\sim 20 \;(100)$~fb at 8 (14)
TeV LHC (the cross section scales as $\tan^{-2}\beta$).

Drell-Yan processes provide another production mode at the LHC.
A recent calculation of the Drell-Yan cross section for $H^+ H^-$,
through a photon or a $Z$, has been provided in Ref.~\cite{Enberg:2013jba}.  As $m_{H^\pm}$ is increased from $\sim 100$~GeV to
$\sim 175$~GeV, $\sigma_{\rm prod}$ decreases from $\sim 50$~fb to $\sim 5$~fb at the 8 TeV LHC; at $14$ TeV the
cross section is roughly twice as large.  We see that the associated $t H^\pm$ production
is a factor of $\sim 4-10$ larger than the Drell-Yan 
cross sections for a similar mass.  (The Drell-Yan and associated 
production only become 
comparable for $\tan\beta$ near 20.  However, note that Drell-Yan 
production has negligible model dependence.)  Hence, single $H^\pm$ production from
$t{\bar t}$ is quite dominant.  However, $H^+H^-$ production from
top pairs, for $\tan\beta=5$, has a cross section $\sim r_t^2 \times 200 {\rm ~ pb}\sim 0.2-20$~fb at 8 TeV,
and is subdominant to Drell-Yan production over most of the relevant parameter space.

Given the large cross section for single production of $H^\pm$ from $t{\bar t}$ pairs,
a careful study of this mode is warranted \cite{HSLee}.  A simple estimate from
our preceding discussion suggests that the 8 TeV LHC data, with $\sim 20$~fb$^{-1}$, contains
$\sim 10^4-10^5$ single charged Higgs events.  The final state here,
$W^+ W^- b\, {\bar b}\, Z_d$, will resemble
that from $t{\bar t}$ production, but with the addition of a $Z_d$.  For $Z_d\to \mu^+\mu^-$,
if the background can be brought under control for $\mu^+\mu^-$
invariant masses below $\sim 10$~GeV,
the existing data would likely probe much of the parameter space.

However, if the $Z_d$ decays invisibly, the search will likely be more challenging and
depend on how well the missing energy signal can be separated from the background.  An approximate
bound on this mode can be inferred from the ATLAS \cite{ATLAS:2013pla} and CMS \cite{Chatrchyan:2013xna}
bounds on stop pair, ${\tilde t}{\tilde t}^*$,
production followed by ${\tilde t}\to t \, \xi^0$, where $\xi^0$ is a neutralino. The final
state here will be $W^+ W^- b \, {\bar b} +$ missing energy, which is the same as
single production of $H^\pm$ (from $t{\bar t}$) 
followed by decay into $W$ and invisible $Z_d$ (though ${\tilde t}{\tilde t}^*$ has 
typically more symmetric missing energy in the event). For neutralino
masses of $\sim 50$~GeV, these experiments find 
bounds of $\sim 2$~pb for a stop mass of $\sim 250$~GeV,
which seems to constrain only the lower part of the mass range
(\ref{massrange}).  We expect that a more detailed analysis, or additional
data from the next run of the LHC, can probe the missing energy signal in
single $H^\pm$ production over a significant part of our parameter space.

In the mass range (\ref{massrange}), most charged Higgs bosons will decay into a
$W$ and a $Z_d$.  The subsequent decay of $Z_d$
will then yield a collimated pair of muons, with a
probability of roughly $20\%$.  The ATLAS collaboration \cite{Aad:2012qua}
has explicitly searched for prompt muon lepton ``jets" (highly collimated lepton pairs)  
in their 2011 data (5 fb$^{-1}$ at 7 TeV).    In one analysis, they search
for two muon ``lepton jets" and the rest of the events are ignored.
An upper bound on the cross section, with mild variations  over the sub-GeV range of $m_{Z_d}$,
of approximately 18 fb is found (we note in passing that, in this ATLAS analysis,
the electron-jet data agree well with expectations, however the muon-jet data
show a mild excess \cite{Aad:2012qua}).

From the preceding discussion, the Drell-Yan $H^+H^-$ cross section $\sigma_{prod}$
will dominate the 2 $Z_d$ production.  We will denote
the branching ratio for $H^+\to W^+ Z_d$, as calculated
above, by $B_{HWZ_d}$.  Also, let $B_{Z_d\mu\mu}$ be the branching fraction for $Z_d\to\mu^+\mu^-$,
assumed to be $20\%$.  The rate for production of $W^+W^-$ and two $\mu^+\mu^-$ pairs is then given by
\beq
\sigma_{\rm prod}\, B_{HWZ_d}^2 B_{Z_d\mu\mu}^2\,.
\label{rate2mu}
\eeq
For $B_{HWZ_d}=1$,
as illustrated above over much of the parameter-space, and for $B_{Z\mu\mu}\sim 0.2$,
we find (for $m_{H^\pm}$ near 100~GeV) that the production rate for $W^+W^-$ with two muon pairs 
is about $\sim 2$~fb.  Even allowing for uncertainties, the aforementioned ATLAS bound \cite{Aad:2012qua}
is still well above the range expected in our framework.  However, the ATLAS analysis ignores
the rest of the event, and the distinct signature in our model contains an additional $W^+ W^-$ pair.
An analysis focused on our particular signature (including its
two-body kinematics), using $\sim 20$~fb$^{-1}$ of 8 TeV data collected in 2012, would presumably
result in an improved bound that will be closer to our expected Drell-Yan signal rate.

Let us now briefly address the regime $m_{H^\pm}\gsim 175$~GeV, where $H^+\to t \, b$ can become
important.  Nonetheless, for $\tan\beta \sim \tan\beta_d$, the $H^\pm \to W^\pm Z_d$
mode will remain significant assuming, as before, 
two body $H^\pm \to W^\pm +$ scalar 
(perhaps except for the SM-like Higgs at 126~GeV) is not kinematically allowed.
Hence, one may typically expect $\ord{1}$ 
branching fraction into $W^\pm \, Z_d$, 
depending on the 2HDM parameters.
In this regime of $m_{H^\pm}$, single
$H^\pm$ production from top pairs, discussed above, will not be important. However,
single production through $b\, g \to H^- \,t$ can be quite important in this case \cite{Klasen:2012wq,Gong:2012ru}.  
The current 8 TeV data could typically contain $\ord{100}$ $b\, g \to H^- \,t$
events and the next run of the LHC (at 14 TeV) can potentially yield several thousand more.
A significant fraction of events in this final state can be expected to yield
$t \, W^-\, Z_d$ which has an extra dilepton or excess missing energy (depending on the $Z_d$ decay), 
compared to single top production in the SM.
In case of $H^- \to {\bar t} \, b$ dominance,
these events contain extra $b$ jets compared to $t {\bar t}$ events, but the large
top pair (including $t{\bar t} \, b {\bar b}$) 
background could pose a challenge for detection of the new physics.

Before closing, we would like to add that there are also $HZZ_d$ and $hZZ_d$ couplings, which will lead to interesting decays of both the light and heavy Higgs, with rates comparable to \eq{HtoWZd-rate}, but depending on scalar mixing angles.   The phenomenology of the $hZZ_d$ decays has been discussed in Ref. \cite{Davoudiasl:2013aya} in the limit in which mixing between the $h$ and $H$ is ignored.    Including the 2 Higgs doublet mixing angle $\alpha$, we have found that the $hZZ_d$ vertex is the same as in Ref. \cite{Davoudiasl:2013aya} where $\cos\beta$ is replaced by $\sin\beta\cos(\beta-\alpha)$.  For large $\tan\beta$ and small $\alpha$, these coincide.   The $HZZ_d$ vertex is given by the $hZZ_d$ coupling with a replacement of $\cos(\beta-\alpha)$ with $-\sin(\beta-\alpha)$ (for
collider phenomenology of alternative dark $Z'$ models, see for example Ref.~\cite{Alves:2013tqa}).

In conclusion, we have considered a type I 2HDM extension of the SM, where one of the doublets
carries a charge under a ``dark" $U(1)_d$ gauge symmetry.  The $U(1)_d$ is well-motivated in this context, as it
forbids large flavor-changing neutral current effects that would violate precision flavor bounds.  The vector boson
associated with this symmetry, $Z_d$, is assumed to be light, and may be relevant in explaining certain
astrophysical signals ascribed to dark matter or the $g_\mu - 2$ anomaly.
We show that in this realization of {\it type I} 2HDM, for
charged Higgs masses in the range $\sim$ 100-175~GeV the decay mode $H^\pm \to W^\pm Z_d$
dominates over the conventional $H^\pm \to \tau^\pm  \nu$ mode, for a wide range of interesting
parameters.  The single production of $H^\pm$, from a top quark decay in a
$t{\bar t}$ final state, is dominant and could potentially cover much
of the parameter space with currently available data, if the $Z_d$ decays into muon pairs $\sim 20\%$ of the time.
In case of an invisibly decaying $Z_d$, the current bounds suggest more data is required to
access significant parts of the parameter space.
For $m_{H^\pm}\gsim 175$~GeV, single production of $H^\pm$ from $t {\bar t}$ pairs is suppressed,
however single production via $b\, g \to H^- \,t$ can provide sizeable statistics.  The next run of
the LHC at 14 TeV can have significant reach for the discovery of $H^\pm$ in the $W^\pm \, Z_d$ mode.

\begin{acknowledgments}

We thank H.-S. Lee for discussions.  The work of H.D. and W.J.M. is supported by the US Department of Energy
under Grant Contract DE-AC02-98CH10886.  The work of R.R. and M.S. is supported by the NSF under Grant No. PHY-1068008.   The opinions and conclusions expressed herein are those of the authors, and do not represent the National Science Foundation.

\end{acknowledgments}

\end{document}